\documentclass[10pt, conference, compsocconf]{IEEEtran}

\usepackage{cite}

\usepackage{color}
\usepackage[table]{xcolor}
\definecolor{lightgray}{gray}{0.9}

\ifCLASSINFOpdf
  \usepackage[pdftex]{graphicx}
\else
  \usepackage[dvips]{graphicx}
\fi
\usepackage{subfig}

\usepackage[cmex10]{amsmath}
\usepackage{chemarr}
\usepackage{amssymb}
\begin{document}
%
\title{Computation of biochemical pathway fluctuations beyond the linear noise approximation using iNA}



%
\author{\IEEEauthorblockN{Philipp
Thomas\IEEEauthorrefmark{1}\IEEEauthorrefmark{2}\IEEEauthorrefmark{3}
\IEEEauthorrefmark{5},
Hannes Matuschek\IEEEauthorrefmark{4}\IEEEauthorrefmark{5} and
Ramon Grima\IEEEauthorrefmark{1}\IEEEauthorrefmark{2}}
\IEEEauthorblockA{\IEEEauthorrefmark{1}School of Biological Sciences, University
of Edinburgh, Edinburgh, United Kingdom}
\IEEEauthorblockA{\IEEEauthorrefmark{2}SynthSys Edinburgh, University of
Edinburgh, Edinburgh, United Kingdom}
\IEEEauthorblockA{\IEEEauthorrefmark{3}Department of Physics, Humboldt
University of Berlin, Berlin, Germany}
\IEEEauthorblockA{\IEEEauthorrefmark{4}Institute of Physics and Astronomy,
University of Potsdam, Potsdam, Germany}
\IEEEauthorblockA{\IEEEauthorrefmark{5}These authors contributed equally.}}


\maketitle

\begin{abstract}
The linear noise approximation is commonly used to obtain intrinsic noise
statistics for biochemical networks. These estimates are accurate for networks 
with large numbers of molecules. However it is well known that many biochemical networks are
characterized by at least one species with a small number of molecules. We here
describe version 0.3 of the software intrinsic Noise Analyzer (iNA) which
allows for accurate computation of noise statistics over wide ranges of molecule
numbers. This is achieved by calculating the next order corrections to the
linear noise approximation's estimates of variance and covariance of
concentration fluctuations. The efficiency of the methods is significantly
improved by automated just-in-time compilation using the LLVM framework leading
to a fluctuation analysis which typically outperforms that obtained by means of
exact stochastic simulations. iNA is hence particularly well suited for the
needs of the computational biology 
community.

\end{abstract}

\begin{IEEEkeywords}
Stochastic modeling; Linear Noise Approximation; genetic regulatory circuits

\end{IEEEkeywords}

%
\IEEEpeerreviewmaketitle

\section{Introduction}
Experimental studies have shown that the protein abundance varies from few tens
to several thousands per protein species per cell \cite{Ishihama2008}. It is
also known that the standard deviation of the concentration fluctuations due to
the random timing of molecular events (intrinsic noise) roughly scales as the
square root of the mean number of molecules \cite{vanKampen}. Hence it is
expected that intrinsic noise plays an important role in the dynamics of those
biochemical networks characterized by at least one species with low molecule
numbers.

The stochastic simulation algorithm (SSA) is the conventional means of probing
stochasticity in biochemical reaction systems \cite{gillespie2007}. This method
simulates every reaction event and hence is typically slow for large reaction
networks; this is particularly true if one is interested in intrinsic noise
statistics which require considerable ensemble averaging of the trajectories
produced by the SSA. A different route of inferring the required statistics
involves finding an approximate solution of the chemical master equation (CME),
a set of differential equations for the probabilities of the states of the
system, which is mathematically equivalent to the SSA. We recently developed
intrinsic Noise Analyzer (iNA) \cite{iNApaper}, the first software package
enabling a fluctuation analysis of biochemical networks via the Linear Noise
Approximation (LNA) and Effective Mesoscopic Rate Equation (EMRE) approximations
of the CME. The former gives the variance and covariance of concentration
fluctuations in the limit of large molecule numbers while the latter gives the mean
concentrations for intermediate to large molecule numbers and is more accurate
than the conventional Rate Equations (REs). 

In this proceeding we develop and efficiently implement in iNA, the Inverse Omega
Square (IOS) method complementing the EMRE method by providing estimates 
for the variance of noise about them. These estimates are accurate for systems whose 
molecule numbers vary over wide ranges (few to thousands of molecules). The new method is tested
on a model of gene expression involving a bimolecular reaction. Remarkably, the
results of the fast IOS calculation are found to agree very well with
those from hour long ensemble averaging over thousands of SSA trajectories.

\section{Results}

In this section we describe the results of the novel IOS method implemented
inside iNA, compare with the results of the RE, LNA and EMRE approximations of
the CME and with exact stochastic simulations using the SSA and finally discuss
the implementation and its performance. The three methods (LNA, EMRE, IOS) are
obtained from the system-size expansion (SSE) of the CME \cite{vanKampen} which
is applicable to monostable chemical systems. Technical details of the various
approximation methods are provided in the section Methods. 

\begin{figure*}
\centering
\includegraphics[width=\textwidth]{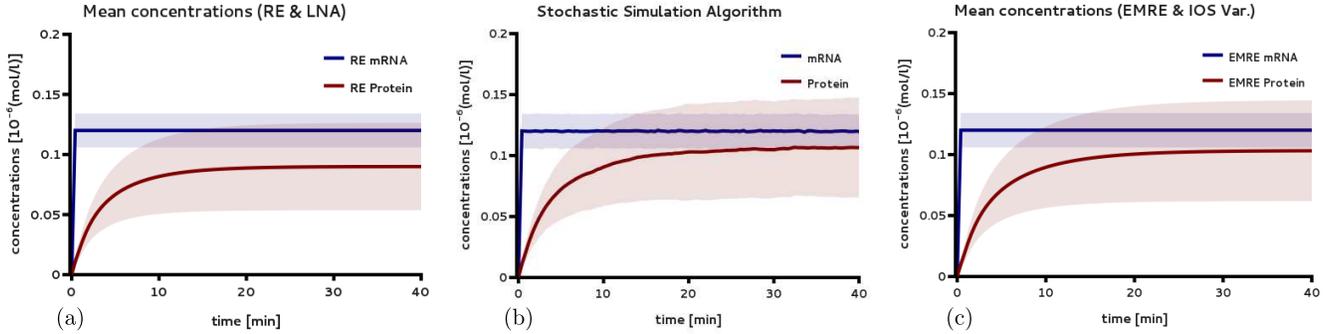}
\caption{Gene expression model with fast transcription rate. In panels (a) and
(b) we compare the RE predictions of mean concentrations with those obtained
from ensemble averaging $3,000$ SSA trajectories. The shaded areas denote the
region of one standard deviation around the average concentrations which in (a)
has been computed using the LNA and in (b) using the SSA. For comparison, in panel (c) we show the
mean concentration prediction according to the EMRE and the variance prediction
according to IOS. The results in (a), (b) and (c) are in good agreement. }
\label{fig:LNA}
\end{figure*}

\subsection{Applications}
We consider a two-stage model of gene expression with an enzymatic degradation
mechanism:
\begin{align}
\label{eqn:gene}
& \text{Gene} \xrightarrow{k_0} \text{Gene} + \text{mRNA}, \ \ \text{mRNA}
\xrightarrow{k_\text{dM}} \varnothing, \notag \\
& \text{mRNA} \xrightarrow{k_s} \text{mRNA} + \text{Protein}, \notag \\
& \text{Protein} + \text{Enzyme} \xrightleftharpoons[k_{-1}]{k_1} \text{Complex}
\xrightarrow{k_2} \text{Enzyme} + \varnothing.
\end{align}
The scheme involves the transcription of mRNA, its translation to protein and
subsequent degradation of both mRNA and protein. Note that while mRNA is
degraded via an unspecific linear reaction, the degradation of protein occurs
via an enzyme catalyzed reaction. The latter may model proteolysis, the
consumption of protein by a metabolic pathway or other post-translational
modifications. A simplified version of this model is one in which the protein
degradation is replaced by the linear reaction: $\text{Protein} \xrightarrow{}
\varnothing$. Over the past decade the latter model has been the subject of
numerous studies, principally because it can be solved exactly since the scheme
is composed of purely first-order reactions
\cite{ozbudak2002,paulsson2005,swain2008}. However, the former model as given by
scheme (\ref{eqn:gene}), cannot be solved exactly because of the bimolecular
association reaction between enzyme and protein. Hence in what follows we
demonstrate the power of approximation methods to infer 
useful information regarding the mechanism's intrinsic noise properties.

{
We consider the model for two parameter sets (see Tab. \ref{tab:params}) at fixed compartmental volume of one femtoliter (one micron cubed). For both cases, the REs predict the same steady state mRNA and
protein concentrations: $[\text{mRNA}]=0.12\mu M$ and $[\text{Protein}]=0.09\mu
M$. These correspond to $72$ and $54$ molecule numbers, respectively. 
}

We have used iNA to compute the mean concentrations using the REs and the
variance of fluctuations using the LNA for parameter set (i) in which
transcription is fast. Comparing these with SSA estimates (see Fig.
\ref{fig:LNA}) we see that the REs and LNA provide reasonably accurate results
for this parameter set. This analysis was within the scope of the previous
version of iNA \cite{iNApaper}. However, the scenario considered is not
particularly realistic. This is since the ratio of protein and mRNA lifetimes in
this example is approximately 100 (as estimated from the time taken for the
concentrations to reach $90\%$ of their steady state values) while an evaluation
of 1,962 genes in budding yeast showed that the ratios have median and mode
close to $3$ \cite{swain2008}. 

\begin{figure*}[t]
\centering
\includegraphics[width=\textwidth]{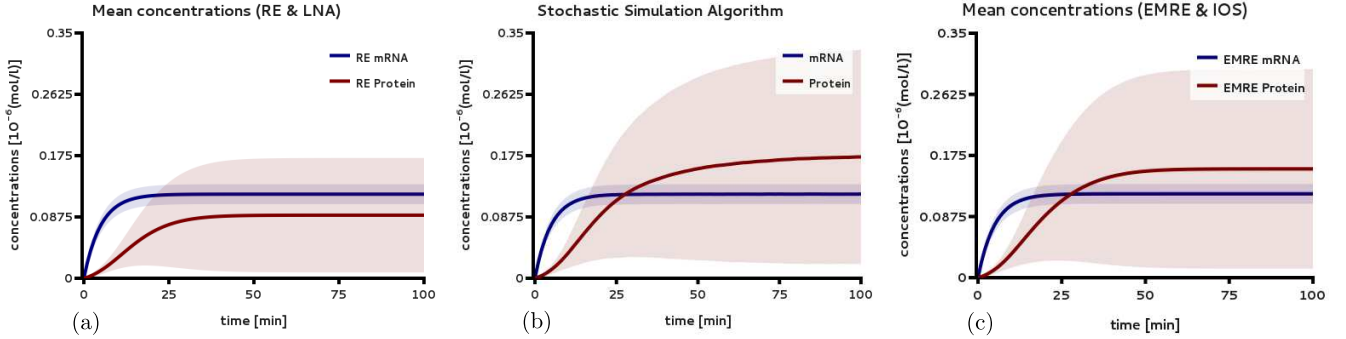}
\caption{Gene expression model with moderate transcription rate. In panels (a)
and (b) we compare the RE and LNA predictions of mean concentrations and
variance of fluctuations with those obtained from ensemble averaging $30,000$
stochastic realizations computed using the SSA. Note that the RE and LNA
predictions are very different than the actual values. In panel (c) we show the
mean concentration prediction according to the EMRE and the variance prediction
according to IOS. These are in good agreement with those obtained from the SSA.}
\label{fig:EMRE}
\end{figure*}

\begin{table}[b]
  \centering
\rowcolors{1}{}{lightgray}
  \begin{tabular}{|l|l|l|}
\hline
    parameter & set (i) & set (ii) \\
\hline
$k_0[G]$ 	& $2.4\text{min}^{-1} \mu M$ 	&  $0.024 \text{min}^{-1} \mu M$ \\
$k_\text{dM}$ 	& $20\text{min}^{-1}$ 		&  $0.2\text{min}^{-1}$		 \\
$k_s$		& $1.5 \text{min}^{-1}$		&  $1.5 \text{min}^{-1}$    	 \\
$k_{-1}, k_{2}$	& $2\text{min}^{-1}$ 		& $2\text{min}^{-1}$		 \\
$k_{1}$ 	& $400(\mu M\text{min})^{-1}$ 	& $400(\mu M\text{min})^{-1}$ \\
    \hline
  \end{tabular}
  \caption{Kinetic parameters used for the gene expression model, scheme
(\ref{eqn:gene}), as discussed in the main text. The volume is fixed to
$\Omega=10^{-15}l$ with an enzyme concentration of $0.1\mu M$ corresponding to
$60$ enzyme molecules. The Michaelis-Menten constant is $0.01 \mu M$ in all
cases.}
  \label{tab:params}
\end{table}

We now consider parameter set (ii), the case of moderate transcription. In Fig.
\ref{fig:EMRE}(a) and (b) we compare the RE and LNA predictions of mean
concentrations and variance of fluctuations with that obtained from the SSA.
Notice that in this case, the two are in severe disagreement. The SSA predicts
that the mean concentration of protein is larger than that of the mRNA while the
REs predict the opposite. This discreteness-induced inversion phenomenon has been described in Ref. \cite{ramaswamy2012}.
It is also the case that the variance estimate of the LNA is considerably smaller than that of the SSA. In Fig. \ref{fig:EMRE}(c) we
show the mean concentrations computed using the EMRE and the variance computed
using the IOS method. Note that the latter are in good agreement with the SSA in
Fig. \ref{fig:EMRE}(b). Note further that while the EMRE was already implemented
in the previous version of iNA, the IOS was not. Hence the present version is
the first to provide estimates of the mean concentrations and of the size of the
noise which are more accurate than both the REs and the LNA. The transient 
times for this case are given by $37$ minutes for protein and $12$ minutes for
mRNA concentrations with a ratio of approximately $3$ in agreement with the
median and mode of experimentally measured ratios. Hence this example provides
clear evidence of the need to go beyond the RE and LNA level of approximation
for physiologically relevant parameters of the gene expression model.

\subsection{Implementation}

iNA's framework consist of three layers of abstraction: the SBML parser which
sets up a mathematical representation of the reaction network, a module which
performs the SSE analytically using the computer algebra system Ginac
\cite{ginac} and a just-in-time (JIT) compiler based on the LLVM \cite{llvm}
framework which compiles the mathematical model into platform specific machine
code at runtime yielding high performance of SSE and SSA analyses implemented in
iNA.
Generally, methods based on the SSE require the numerical solution of a high dimensional system of linear equations. In the present version 0.3 of iNA, time-course analysis is efficiently performed by means of the all-round ODE integrator LSODA which automatically switches between explicit and implicit methods \cite{petzold1983}.
%
%
In particular, the number of simultaneous equations to be solved for the LNA, EMRE analysis is
approximately proportional to $N^2$ where $N$ is the number of independent
species after conservation analysis \cite{vallabhajosyula2006} while for IOS,
the number of equations is approximately proportional to $N^3$. Quadratic
and cubic dependencies represent a challenge for software development. iNA's
previous version 0.2 addressed this need for performance by providing a bytecode interpreter (BCI) 
for efficient expression evaluation \cite{iNApaper}. The latter concept has proven its performance for
both SSE and SSA methods while maintaining compatibility over many platforms.
With the present version we introduce a strategy for JIT compilation based on the modern compiler framework
LLVM that allows to emit and execute platform specific machine code at runtime \cite{llvm}. The system size expansion ODEs are automatically compiled as native machine code making it executable directly on the CPU. Therefore iNA's JIT feature enjoys the speed of statically compiled code while maintaining the flexibility common to interpreters. 
Moreover, the LLVM framework allows optimizations on platform independent and platform specific instruction levels  which are beneficial for computationally expensive calculations.

Compared to iNA's previous implementation, using JIT compilation, we have observed speedups for the SSE analysis by factors of $10-20$ and factors of $1.5-2$ for the SSA.  

\section{Methods}

We consider a general reaction network confined in a volume $\Omega$ under
well-mixed conditions and involving the interaction of $N$ distinct chemical
species via $R$ chemical reactions of the type
 \begin{equation}
\label{eqn:reaction}
 s_{1j} X_1 + \ldots + s_{Nj} X_{N} \xrightarrow{k_j} \ r_{1j} X_{1} + \ldots +
r_{Nj} X_{N},
 \end{equation}
where $j$ is the reaction index running from $1$ to $R$, $X_i$ denotes chemical
species $i$, $k_j$ is the reaction rate of the $j^{th}$ reaction and $s_{ij}$
and $r_{ij}$ are the stoichiometric coefficients. 
Note that our general formulation does not require all reactions to be
necessarily elementary. 

The CME gives the time-evolution equation for the probability $P(\vec{n},t)$
that the system is in a particular mesoscopic state $\vec{n}=(n_1,...,n_N)^T$
where  $n_i$ is the number of molecules of the $i^{th}$ species. It is given by:
 \begin{equation}
 \label{eqn:CME}
 \frac{\partial P(\vec{n},t)}{\partial t} = \sum_{j=1}^{R} \biggl(
\displaystyle\prod_{i=1}^N E_i^{-S_{ij}} - 1 \biggr)
\hat{a}_j\left(\vec{n},\Omega\right) P(\vec{n},t),
 \end{equation}
where $S_{ij} = r_{ij}-s_{ij}$, $\hat{a}_j(\vec{n},\vec\Omega)$ is the
propensity function such that the probability for the $j^{th}$ reaction to occur
in the time interval $[t,t+dt)$ is given by $\hat{a}_j(\vec{n},\vec\Omega) dt$
\cite{gillespie2007} and  $E_i^{-S_{ij}}$ is the step operator defined by its
action on a general function of molecular populations as $
E_i^{-S_{ij}}g(n_1,...,n_i,...,n_N)=g(n_1,...,n_i-S_{ij},...,n_N)$
\cite{vanKampen}.

The CME is typically intractable for computational purposes because of the
inherently large state space. iNA circumvents this problem by approximating the
moments of probability density solution of the CME systematically using van
Kampen's SSE \cite{vanKampen,vanKampen1976}. The starting point of the analysis
is van Kampen's ansatz
\begin{align}
 \label{eqn:cov}
 \frac{\vec{n}}{\Omega}=[\vec{X}] + \Omega^{-1/2}\vec\epsilon,
\end{align}
by which one separates the instantaneous concentration into a deterministic part
given by the solution $[\vec{X}]$ of the macroscopic REs for the reaction scheme
(\ref{eqn:reaction}) and the fluctuations around it parametrized by
$\vec\epsilon$. The change of variable causes the probability distribution of molecular
populations $P(\vec{n},t)$ to be transformed into a new probability distribution
of fluctuations $\Pi(\vec\epsilon,t)$. The time derivative, the step operator
and the propensity functions are also transformed (see \cite{iNApaper} for
details). In particular, using ansatz (\ref{eqn:cov}) together with the explicit $\Omega$-dependence of the propensities as given by ${\hat{a}_j(\vec{n},\Omega)}=\sum_{m=0}^\infty \Omega^{1-m} f_j^{(m)}(\frac{\vec{n}}{\Omega})$, the propensities can be expanded in powers of $\Omega^{-1/2}$:
\begin{align}
\label{eqn:prop}
 \frac{\hat{a}_j(\vec{n},\Omega)}{\Omega}=&  
f_j^{(0)}([\vec{X}]) +
 \Omega^{-1/2}\epsilon_\alpha \frac{\partial
f_j^{(0)}([\vec{X}])}{\partial{[{X}_\alpha]}} + \Omega^{-1}f_j^{(1)}([\vec{X}])
\notag \\
 &+ \frac{1}{2}\Omega^{-1}\epsilon_\alpha \epsilon_\beta \frac{\partial
f_j^{(0)}([\vec{X}])}{\partial{[X_\alpha]}\partial[X_\beta]} +
 \Omega^{-3/2}\epsilon_\alpha \frac{\partial
f_j^{(1)}([\vec{X}])}{\partial{[{X}_\alpha]}}
\notag \\
 & + \frac{1}{2}\Omega^{-3/2}\epsilon_\alpha \epsilon_\beta \epsilon_\gamma \frac{\partial
f_j^{(0)}([\vec{X}])}{\partial{[X_\alpha]}\partial[X_\beta]\partial{[X_\gamma]}}  
 + O(\Omega^{-2}).
\end{align}
Note that $\vec{f}^{(0)}$ is the macroscopic rate function. Here we have used the convention that twice repeated Greek indices are summed over 1 to $N$, which we use in what follows as well. Consequently the CME
up to order $\Omega^{-1}$ can be written as
\begin{align}
\label{eqn:cmeexp}
 &\frac{\partial \Pi(\vec{\epsilon},t)}{\partial t}-\Omega^{1/2}
 \left(\frac{\partial [X_\alpha]}{\partial t}-\sum_{k=1}^{R} S_{\alpha k}
f_k^{(0)}([\vec{X}]) \right) {\partial_\alpha  \Pi(\vec{\epsilon},t)} 
\notag\\
&\qquad = \left( \Omega^{0} \mathcal{L}^{(0)} 
 + \Omega^{-1/2} \mathcal{L}^{(1)} + \Omega^{-1} \mathcal{L}^{(2)}\right)\,
\Pi(\vec{\epsilon},t) \notag\\ &\qquad \qquad  + O(\Omega^{-3/2}),
\end{align}
where the operators are defined as
\begin{align}
\label{eqn:vertices}
\mathcal{L}^{(0)}&=
 -\partial_\alpha
{J}_{\alpha}^{\beta}\epsilon_\beta+\frac{1}{2}\partial_\alpha\partial_\beta
D_{\alpha\beta}, \notag
\\
\mathcal{L}^{(1)}&=
 -\partial_\alpha {D}_{\alpha}^{(1)}
-\frac{1}{2!} \partial_\alpha{J}_{\alpha}^{\beta\gamma}  \epsilon_\beta
\epsilon_\gamma 
+ \frac{1}{2!} \partial_\alpha \partial_\beta{J}_{\alpha\beta}^\gamma 
\epsilon_\gamma \notag \\ 
&-  \frac{1}{3!}\partial_\alpha \partial_\beta \partial_\gamma {D}_{\alpha \beta
\gamma},\notag
\\
\mathcal{L}^{(2)}&= -\partial_\alpha J^{(1)\beta}_{\alpha}  \epsilon_\beta 
+ \frac{1}{2!} \partial_\alpha \partial_\beta {D}^{(1)}_{\alpha\beta} 
- \frac{1}{3!} \partial_\alpha  {J}_{\alpha}^{\beta\gamma\delta}\epsilon_\beta
\epsilon_\gamma \epsilon_\delta \notag \\ 
&+ \frac{1}{2!}\frac{1}{2!} \partial_\alpha \partial_\beta 
{J}_{\alpha\beta}^{\gamma\delta}\epsilon_\gamma \epsilon_\delta
-  \frac{1}{3!}  \partial_\alpha \partial_\beta
\partial_\gamma{J}_{\alpha\beta\gamma}^{\delta} \epsilon_\delta \notag \\ 
 &+ \frac{1}{4!} \partial_\alpha \partial_\beta \partial_\gamma
\partial_\delta{D}_{\alpha \beta \gamma \delta},
\end{align} 
and the SSE coefficients are given by
\begin{align}
{D^{(n)}}_{ij..r}&=\sum\limits_{k=1}^R
S_{ik}S_{jk}...S_{rk}\,{f}^{(n)}_k([\vec{X}]),\notag\\
{J^{(n)}}_{ij..r}^{st..z}&=\frac{\partial}{\partial{[X_s]}}\frac{\partial}{
\partial{[X_t]}}...\frac{\partial}{\partial{[X_z]}} {D^{(n)}}_{ij..r}.
\end{align}
Note that the above expressions generalize the expansion carried out in Ref.
\cite{grima2011} to include also non-elementary reactions as for instance
trimolecular reactions or reactions with propensities of the Michaelis-Menten
type \cite{rao2003}. Note also that the $\Omega^{1/2}$ term vanishes since the
macroscopic REs are given by $\partial_t [X_\alpha]=\sum_{k=1}^{R} S_{\alpha k}
f_k^{(0)}([\vec{X}])$ leaving us with a series expansion of the CME in powers of
the inverse square root of the volume. In Eqs. (\ref{eqn:vertices}) we have omitted the upper
index in the bracket of the SSE coefficients in the case of $n=0$.

The method now proceeds by constructing equations for the moments of the $\vec\epsilon$
variables. This is accomplished by expanding the probability distribution of fluctuations $\Pi(\vec{\epsilon},t)$ in
terms of the inverse square root of the volume as
$\Pi(\vec{\epsilon},t)=\sum_{j=0}^\infty \Pi_j(\vec{\epsilon},t)\Omega^{-j/2}$,
which implies an equivalent expansion of the moments
\begin{align}
 \label{eqn:mExp}
 \langle \epsilon_k\epsilon_l ...\epsilon_m \rangle=\sum_{j=0}^\infty
[\epsilon_k\epsilon_l ...\epsilon_m]_j\Omega^{-j/2}.
\end{align}
In order to relate the above moments back to the moments of the concentration
variables we use Eqs. (\ref{eqn:cov}) and (\ref{eqn:mExp}) to find expressions
for the mean concentrations and covariance of fluctuations which are given by
\begin{subequations}
\begin{align}
 \label{eqn:defMean}
 \left\langle \frac{n_i}{\Omega}\right\rangle &= \, [X_i] +
{\Omega}^{-1}\left[\epsilon_i\right]_1 +O(\Omega^{-2}),
\\
 \Sigma_{ij} &= \,\left\langle
                \left( \frac{n_i}{\Omega}-\left\langle
\frac{n_i}{\Omega}\right\rangle\right)\left(\frac{n_j}{\Omega}-\left\langle
\frac{n_j}{\Omega}\right\rangle \right) \right\rangle \notag \\ 
&= \, \Omega^{-1}
\left[\epsilon_i\epsilon_j\right]_0 + \Omega^{-2}\left(\left[\epsilon_i\epsilon_j\right]_2
-\left[\epsilon_i\right]_1\left[\epsilon_j\right]_1\right)\notag
 \\& \qquad+ O(\Omega^{-3})  \label{eqn:defSigma}
\end{align}
\end{subequations}
The order $\Omega^0$ term of Eq. (\ref{eqn:defMean}) denotes the mean concentrations as given by the macroscopic REs while the $\Omega^{-1}$ term in Eq. (\ref{eqn:defSigma}) gives the LNA estimate for the covariance.
Including terms to order $\Omega^{-1}$ in Eq. (\ref{eqn:defMean}) gives the EMRE estimate of the mean concentrations which corrects the estimate of the REs. Finally, considering also the $\Omega^{-2}$ term in Eq. (\ref{eqn:defSigma}) gives the IOS (Inverse Omega Squared) estimate of the variance which is centered around the EMRE concentrations and is of higher accuracy than the LNA method. The general procedure together with the equations determining the coefficients of Eq. (\ref{eqn:mExp}) is presented in Refs. \cite{grima2011} and \cite{grima2012}. In brief the result can be summarized as follows: as mentioned above, truncating Eq. (\ref{eqn:cmeexp}) to order $\Omega^{1/2}$ yields the macroscopic REs, truncation after terms up to order $\Omega^{0}$ gives a Fokker-Planck equation with linear drift and diffusion coefficients which is also called the Linear Noise Approximation. Considering terms up to $\Omega^{-1/2}$ one obtains the EMREs while the terms up to $\Omega^{-1}$ determine the corrections beyond 
the LNA as given by the 
present IOS method.
Since the IOS method has been derived from van Kampen's ansatz, Eq. (\ref{eqn:cov}), which expands the CME around the macroscopic concentrations it has the same limitation as the LNA namely that it cannot account for systems exhibiting bistability.
  
\section{Discussion}
In this proceeding we have introduced and implemented the IOS approximation in the
software package iNA. {This allows the variances to be
determined accurate to order $\Omega^{-2}$, an approximation which complements the EMRE method (mean concentrations accurate to order $\Omega^{-1}$) and is superior to the previously implemented LNA method (variances accurate to order
$\Omega^0$).} As we have shown this increased accuracy is desirable to accurately account for the
effects of intrinsic noise in biochemical reaction networks under low molecule
number conditions. In particular, we have demonstrated the utility of the
software by analyzing an example of gene expression with a functional enzyme. We have also extended iNA by a more efficient JIT compilation strategy in combination with improved numerical algorithms which offers high performance
and enables computations feasible even on desktop PCs. This feature is
particularly important when analyzing noise in reaction networks of intermediate or large
size with bimolecular reactions, conditions that have been shown to amplify the deviations from the conventional rate
equation description \cite{thomas2010}.


\section*{Acknowledgment}
RG gratefully acknowledges support by SULSA (Scottish Universities Life Science
Alliance).

\section*{Availability}
The software iNA version 0.3 is freely available under
http://code.google.com/p/intrinsic-noise-analyzer/ as executable binaries for
Linux, MacOSX and Microsoft Windows, as well as the full source code under an
open source license.



\bibliographystyle{unsrt}
\bibliography{ina2}

\end{document}